# Identification of vacancy defects in a thin film perovskite oxide

D.J. Keeble,[1,*] R.A. Mackie,[1] W. Egger,[2] B. Löwe,[2] P. Pikart,[3] C. Hugenschmidt,[3] and T.J. Jackson[4]

[1]*Carnegie Laboratory of Physics, School of Engineering, Physics, and Mathematics, University of Dundee, Dundee DD1 4HN, United Kingdom*

[2]*Universität Bundeswehr München, D-85577 Neubiberg, Germany*

[3]*Technische Universität München, ZWEFRM 11 E21, D-85747 Garching, Germany*

[4]*Department of Electronic, Electrical and Computer Engineering, University of Birmingham, Birmingham B15 2TT, United Kingdom*

[*] Electronic address: d.j.keeble@dundee.ac.uk

Vacancy defects in thin film laser ablated $SrTiO_3$ on $SrTiO_3$ were identified using variable energy positron annihilation lifetime measurements. Strontium vacancy related defects were the dominant positron traps and, apart from in the top ~ 50 nm, were found to be uniformly distributed. The surface layer showed an increase in annihilation from larger open-volume defects, large vacancy clusters or nanovoids.

The ability to produce high quality perovskite oxide, $ABO_3$, thin films and heterostructures has led the development of oxide electronics [1-4] and to an increase in our understanding of ferroic and multiferroic materials [5, 6]. The marked improvements in the structural properties of perovskite oxide layers will ultimately result in the electronic properties being limited by the presence of electrically active point defects at sub-ppm concentration levels. Ferroic perovskite oxides are typically wide band-gap semiconductors with $E_g \sim$ 3.5 to 4.1 eV [1]. Point defects in ferroics can alter material properties in two ways; they act, as in conventional semiconductors, as charge sources, sinks, and recombination centers, but in addition the strain and electric fields associated with the defect can interact directly with the mesoscopic polarization and, for example, pin domain walls [7].

Vacancy defect complexes with a net local dipole moment are of particular relevance. Electron magnetic resonance methods have the required sensitivity and can provided detailed local structure models for those centers with a paramagnetic ground state [8]. However, convincing evidence is lacking that the isolated native vacancy defects have accessible EPR active states in perovskite oxides [9]. Direct imaging of O vacancies has been demonstrated using aberration corrected transmission electron microscopy (TEM) [10], or annular-dark-field TEM, supported by electron energy-loss spectroscopy (EELS) [11]. Electron energy-loss near-edge structures (ELNES) combined with first-principles calculations provided evidence for high concentrations of Sr vacancies in the vicinity of grain boundaries, resulting from heat treatments [12]. However, these methods typically require defect concentrations greater than 1%, and are time consuming and destructive.

Here we report the detection of a Sr vacancy related defect distribution in a thin film of the perovskite oxide $SrTiO_3$ grown by laser ablation using positron lifetime measurements performed with a an ultra-high intensity variable energy positron beam. The dominant positron lifetime component in the thin film at ~260 ps is due to trapping at Sr vacancy related defects. Vacancy cluster defects with a lifetime of ~ 410 ps were also detected. A near-surface, ~ 50 nm, layer was resolved and showed an increase in trapping to large vacancy clusters or nanovoids. Previous positron annihilation studies of oxide thin films have typically been performed using conventional positron beams with 2-3 orders of magnitude lower intensity, and have used Doppler broadening spectroscopy which gives less direct information on the nature of the trapping vacancy.

An implanted positron will thermalize within a few picoseconds then annihilate in the material from a state $i$ with a lifetime $\tau_i$ and a probability $I_i$. This can be a delocalized state in the bulk lattice, or a localized state at a vacancy defect, or if the open-volume defect has sufficient size (*e.g.* in nanovoids) the electron-positron bound state known as positronium can be formed. If the average lifetime, $\bar{\tau} = \sum_i I_i \tau_i$, is greater than the bulk lattice lifetime characteristic of the material, $\tau_B$, this indicates that vacancy-type defects are present. The rate of positron trapping to a vacancy, $\kappa_d$, is proportional to the concentration of these defects, $[d]$, where the constant of proportionality is the defect specific trapping coefficient, $\mu_d$; $\kappa_d = \mu_d [d]$. While vacancy defects, due to the lack of a positive ion core, form an attractive potential the local charge must also be considered, if positive a Coulomb barrier inhibits trapping, however, for neutral or negative defects trapping coefficients are large, typically in the range $0.5 \times 10^{15}$ to $6 \times 10^{15}$ s$^{-1}$ at., giving the method sub-ppm concentration sensitivity [13].

The one defect simple trapping model (1D-STM) predicts two experimental positron lifetimes; $\tau_2 = \tau_d$ is the lifetime characteristic of the defect, and the first lifetime, $\tau_1$, is reduced from the bulk, or perfect lattice, lifetime, $\tau_B$, by an amount that depends on the rate of trapping to the defect, so that $\tau_1 \leq \tau_B$. The model is readily extended to two or more defects, and allows defect positron trapping rates, $\kappa_d$, to be determined [13]. As the concentration of vacancies increases, the intensity of the longer defect lifetime, $I_2$, increases towards unity. Eventually saturation trapping occurs where all positrons annihilate from the defect and sensitivity to concentration is restricted or lost completely; assuming a specific trapping coefficient $\sim 2 \times 10^{15}$ s$^{-1}$ at. and $\tau_B = 150$ ps this would occur for a concentration of the order of 50 ppm.

The positron lifetimes for perfect material and for specific types of vacancy defect can be calculated using density functional theory (DFT) methods; the values obtained for SrTiO$_3$ are given in Table I [14]. The lifetimes for positrons trapped at localized states in cation vacancies are significantly larger than the $\sim 150$ ps obtained for the perfect lattice, bulk, state. Experimental lifetimes of $\sim 181$ ps for the Ti vacancy and $\sim 275$ ps for Sr vacancy related defects have been inferred from studies of single crystal SrTiO$_3$, and along with the bulk lifetime, are consistent with the DFT values [14].

Measurements were performed on a thin film of $SrTiO_3$ deposited by pulsed laser ablation on $SrTiO_3$ [15]. The thickness was determined to be 603(6) nm by TEM. A KrF excimer laser was used, giving a fluence at the target surface of 1.5 J cm$^{-2}$, 5000 pulses at a repetition rate of 0.15 Hz, the film was grown using 5000 pulses with a temperature 750 ºC and the oxygen partial pressure was 40 Pa. Double side polished (100) single crystal $SrTiO_3$, supplied by MaTecK GmbH, was also studied. Positron annihilation measurements were performed at the ultra-high intensity neutron induced positron source (NEPOMUC) at the Munich research reactor FRMII with a primary moderated beam intensity of $5\times10^8$ e$^+$ s$^{-1}$ at an energy of 1 keV [16]. Variable energy (VE) positron annihilation lifetime spectroscopy (PALS) measurements were made with the pulsed beam instrument comprising a pre-buncher, chopper, and main buncher operating at 50 MHz providing the time structure and start timing signal, the annihilation radiation from the implanted positrons was detected using $BaF_2$ scintillation detector [17]. The beam energy could be varied between 0.5 to 22 keV. The lifetime spectra contained an average of $5.8\times10^6$ counts accumulated with a count rate of $\sim6\times10^3$ s$^{-1}$, the instrument timing resolution function was normally described by three dominant, energy dependent, terms; these showed a mean width, averaged over all energies, of 306 ps. In addition, measurements were made with the beam transported to the Doppler broadening spectroscopy (DBS) instrument [18].

The mean positron lifetime as a function of positron implantation energy for the laser ablated $SrTiO_3$ on $SrTiO_3$ sample is shown in Fig. 1(a). This deceases from ~350 ps at the surface rapidly to ~280 ps and remains at this value through the bulk of the film. The onset of a reduction from ~280 ps occurs at ~9 keV, from which point the broadened Makhovian implantation profiles increasingly penetrate beyond the film-substrate interface at ~600 nm and sample the substrate, as shown in Fig. 1(e). From the large value of the mean lifetime, $\bar{\tau}$, ~280 ps in the film, knowing $\tau_B$, and considering a plausible range of defect specific trapping coefficients, it can be concluded that the positrons are saturation trapped at vacancy defects.

Figure 1(c) shows measurements on the single crystal $SrTiO_3$ substrate, the mean lifetime, $\bar{\tau}$, decreasing rapidly from a near-surface value of 266 ps to ~170 ps. The $\bar{\tau}$ obtained by conventional PALS [14], using unmoderated positrons implanting to a continuous distribution of depths down to ~0.2 mm, was 167 ps.

The rapidly accumulated high number of counts, the well-defined instrument timing resolution, and the lack of positron source annihilation terms, allowed reliable multi-exponential de-convolution of the spectra. The results for the $SrTiO_3$ on $SrTiO_3$ thin film and the $SrTiO_3$ single crystal are shown in Fig. 2. The average fit chi-squared values were 1.15(9) and 1.16(10), respectively, and three lifetime components were resolved. The variation of lifetime values and intensities with implantation depth for the laser ablated film clearly show a ~ 50 nm surface region below which the vacancy defect densities remain constant through the film, the interface with the substrate is then clearly observed. Positrons implanted with energies greater than ~ 9 keV increasingly sample the film interface and substrate. At the highest implantation energies the dominant lifetime values are tending to those observed below the near-surface region of the single crystal sample.

The top layer of the SrTiO$_3$ film is characterized by observation of long third lifetime component with a value decreasing from ~ 2 to 3 ns from the surface with a concomitant decrease in intensity, Fig. 2(c) and (d). This is consistent with pick-off annihilation events due to the formation of orthopositronium in large vacancy clusters or nanovoids, and with a possible contribution from back diffusion to the surface followed by positronium formation [13]. The intensity of the ~ 320-400 ps component in the near-surface layer (Fig. 2(a) and (b)) indicates the presence of larger vacancy cluster defects.

The uniform values for the lifetime and intensities of the resolved components the SrTiO$_3$ film were observed below this surface layer for implantation energies between 4–8 keV; the two main components were $\tau_1$ = 257(1) ps, I$_1$ = 87(1) %, and $\tau_2$ = 407(10) ps, I$_2$ = 13(1) %. A third nanosecond component had an intensity of 0.26(3) % through this energy range and has been omitted from Fig. 2(a-d) for clarity.

The dominant lifetime component, ~257 ps (~87 %), in the main extent of the thin film has a value slightly less than that determined for V$_{Sr}$, ~275 ps. A similar ~260 ps lifetime was observed for electron irradiated SrTiO$_3$ and vacuum annealed SrTiO$_3$ where the pre-treatment samples showed a component attributable to Ti vacancies [14]. Saturation trapping to both Ti (~181 ps) and Sr (~275 ps) vacancies result, due to the finite instrument resolution function, in a single intermediate value lifetime component, a weighted average determined by the ratio of the vacancy concentrations and the defect specific trapping coefficients for the two defects ($\mu_{VTi}[V_{Ti}]/\mu_{VSr}[V_{Sr}]$). The defect specific trapping rate for the B-site vacancy is greater than that for A-site [19], assuming they differ by a factor 2 then simulations, similar to those described in Ref. [14], show $[V_{Sr}] > 4[V_{Ti}]$ would be required to obtain a component lifetime close to 260 ps [14]. The differences in the lifetimes between isolated Sr monovancancies, and $V_{Sr}$-$nV_O$ complexes, see Table I, are small and cannot be reliably differentiated.

The low intensity second lifetime component, 407(10) ps, observed for the same implantation energies (Fig. 2(a,b)) is due to open volume defects larger than monovacancy, or cation vacancy nearest neighbor oxygen vacancy complexes (Table I). Here the MIKA-DFT calculations have been extended, using 1080 atom supercells, to cation divacancy complexes, see Table I. Comparable lifetime values were obtained for Sr divacancy defects with or without oxygen vacancy nearest neighbors of ~ 283 ps, similarly linear configuration Ti divacancy defects with between one and three oxygen vacancy nearest neighbors gave lifetime values between 247 and 253 ps, respectively. The largest value, 316 ps, was obtained for a Ti-Sr divacancy with three nearest neighbor oxygen vacancies. These calculations provide evidence that the experimental lifetime corresponds to a larger vacancy cluster defect.

The component positron lifetimes for the SrTiO$_3$ single crystal substrate are shown in Fig. 2(e-h). Three lifetime components are required to fit the data reliably; a low intensity, 1.4(5) %, 1-2 ns orthopositronium pick-off third lifetime, a dominant short lifetime varying from ~115 ps at 0.5 keV near the surface to ~140 ps at 18 keV (corresponding to a mean implantation depth of ~750 nm) and a second lifetime varying from 403 ps to 213 ps, with increasing intensity, over the same implantation range. The PALS spectra obtained using un-moderated positrons from the sample were fitted using two lifetime components with a chi-squared of 1.009 and gave $\tau_1$ = 123(3) ps, $I_1$ = 45(4) %, and $\tau_2$ = 202(3) ps, with appropriate resolution function and source correction terms included.

These results provide evidence that the polished crystal contains larger open-volume defects giving rise to both the nanosecond component and the ~400 ps component, and that these are mainly confined to the top few hundred nanometers. However, in contrast to the thin film, a short reduced bulk positron lifetime component is clearly resolved and is due to a fraction of annihilation events coming from perfect lattice. The lifetimes detected at 18 keV, and using conventional PALS, are in agreement with previous studies on as-received $SrTiO_3$ crystal samples [14], the ~200 ps component can be attributed to unresolved contributions from $V_{Ti}$ and $V_{Sr}$ defects with concentrations in the ~0.1–1 ppm range.

Further insight can be obtained on the interface of the $SrTiO_3$ thin film with the substrate. The variation in the lifetime components for energies greater than 8 keV in the thin film $SrTiO_3$ sample clearly show the onset of annihilation events from the film interface and the substrate (Fig. 2(a-b)). A low intensity third lifetime component due to large open-volume defects is detected, and a short reduced bulk lifetime is resolved, the dominant second lifetime reduces from ~400 ps to ~270 ps between 8-12 keV. For the implantation energies 13–18 keV the average values for the second and third lifetimes are 235(13) ps and 480(70) ps, respectively, these are comparable to the results from near-surface region of the as-received substrate (Fig. 2(e-h)). However, the intensity of the dominant defect lifetime is higher consistent with near surface defect formation in the substrate during deposition.

Figure 1 also shows the Doppler broadening spectroscopy detected *S* parameter depth profiles for the SrTiO$_3$ thin film and single crystal SrTiO$_3$ samples obtained immediately after the VE-PALS measurements. This parameter describes the shape of the annihilation radiation energy spectrum. It is a measure of the fraction of positrons annihilating mainly with valence electrons with low longitudinal momentum component values, $|p_L| < 3.1 \times 10^{-3} m_0 c$. Positron trapping to vacancy-related defects results in an increase in *S* parameter compared to perfect material. The magnitude of the increase can provide insight on the size of the dominant open-volume defect, and on defect density. The increase in the *S* between the substrate value and the plateau in the thin film (5–10 keV) is ~ 3.9 %, it has been suggested that this is consistent with trapping to Sr vacancies or V$_{Sr}$-$n$V$_O$ complexes [20]. The further marked increase in *S* toward the surface of film, and the substrate, is often characteristic of the presence of larger open-volume defects. Comparison of Fig. 1(b) and (d) with behavior of the positron lifetime components shown in Fig. 2 provide direct evidence for validity of these assumptions.

In conclusion, depth dependent positron lifetime measurements performed using an ultra-high intensity positron beam show that Sr vacancy related defects are the dominant positron traps in the laser ablated SrTiO$_3$ on SrTiO$_3$ thin films studied, and below a near surface region of depth ~50 nm have a uniform distribution. A second longer lifetime component was also observed and is attributed to vacancy cluster defects larger than cation divacancy nearest neighbor oxygen vacancy complexes.

RAM acknowledges the support from the Carnegie Trust for the Universities of Scotland. DJK and RAM acknowledge support of the European Commission Programme RII3-CT-2003-505925.

13## Tables and Table Captions

TABLE I. Positron lifetime values (ps) calculated by the DFT method MIKA for monovacancy and vacancy nearest neighbor complexes in SrTiO$_3$, (relaxed structure calculations in parentheses). LMTO DFT method values from Ref. [14] are also shown.

|         | Bulk | $V_O$ | $V_{O-O}$ | $V_{Ti}$ | $V_{Ti-O}$ | $V_{Sr}$ | $V_{Sr-O}$ | $V_{Sr-3O}$ | $V_{Sr-Sr}$ | $V_{Ti-O-Ti}$ | $V_{Ti-3O-Sr}$ |
|---------|------|-------|-----------|----------|------------|----------|------------|-------------|-------------|---------------|----------------|
| MIKA-AP | 151  | 166   | 178       | 195      | 225        | 279      | 283        | 289         | 283         | 247           | 316            |
|         |      | (170) |           | (184)    |            | (279)    |            |             |             |               |                |
| LMTO-AP | 146  | 159   |           | 194      |            | 252      |            |             |             |               |                |

**Figure Captions**

FIG. 1. (Color online) Mean positron lifetimes against positron implantation energy, measured by variable energy positron beam positron annihilation lifetime spectroscopy, for (a) $SrTiO_3$ thin film on $SrTiO_3$, and (c) $SrTiO_3$ single crystal substrate. The Doppler broadening spectroscopy detected low momentum fraction *S* parameter depth profiles for the two samples are shown in (b) and (d), respectively. The Makhovian positron implantation profiles calculated for $SrTiO_3$ are shown in (e).

FIG. 2. (Color online) The results of multi-exponential positron lifetime deconvolution for $SrTiO_3$ thin film on $SrTiO_3$ (a-d), and $SrTiO_3$ single crystal substrate (e-h).

FIG. 1

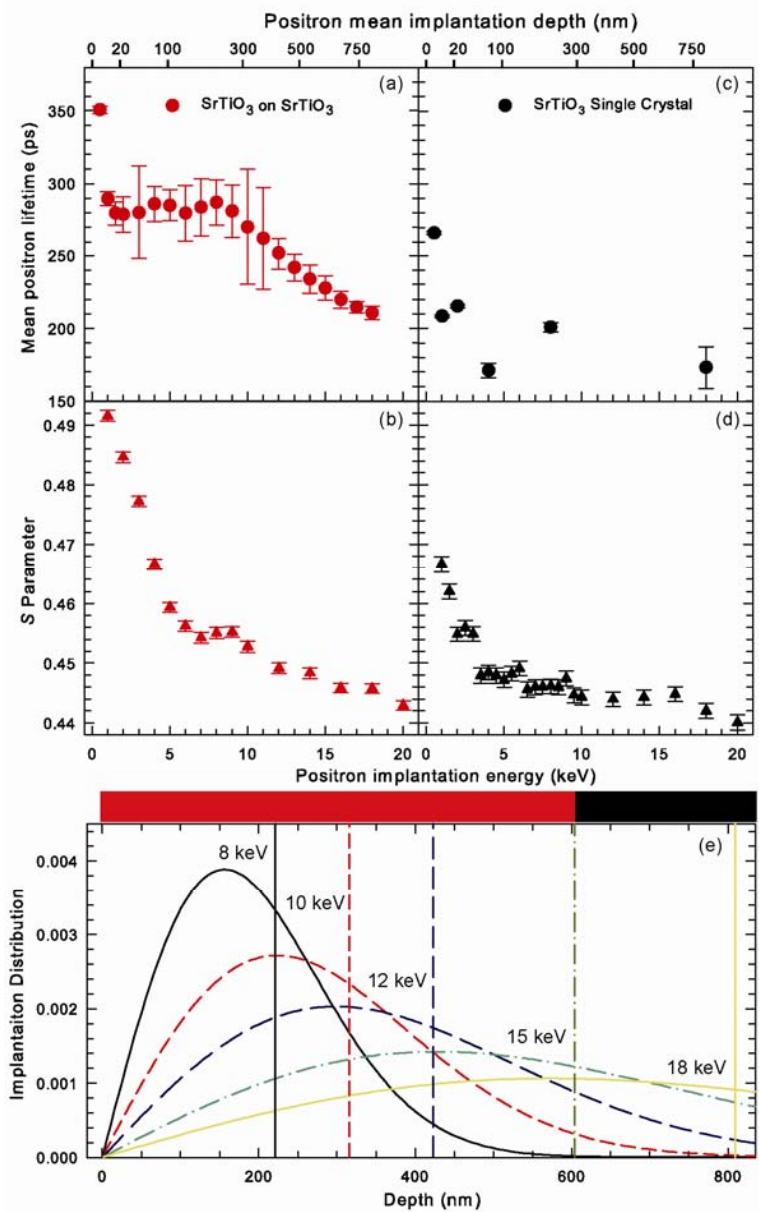

FIG. 2

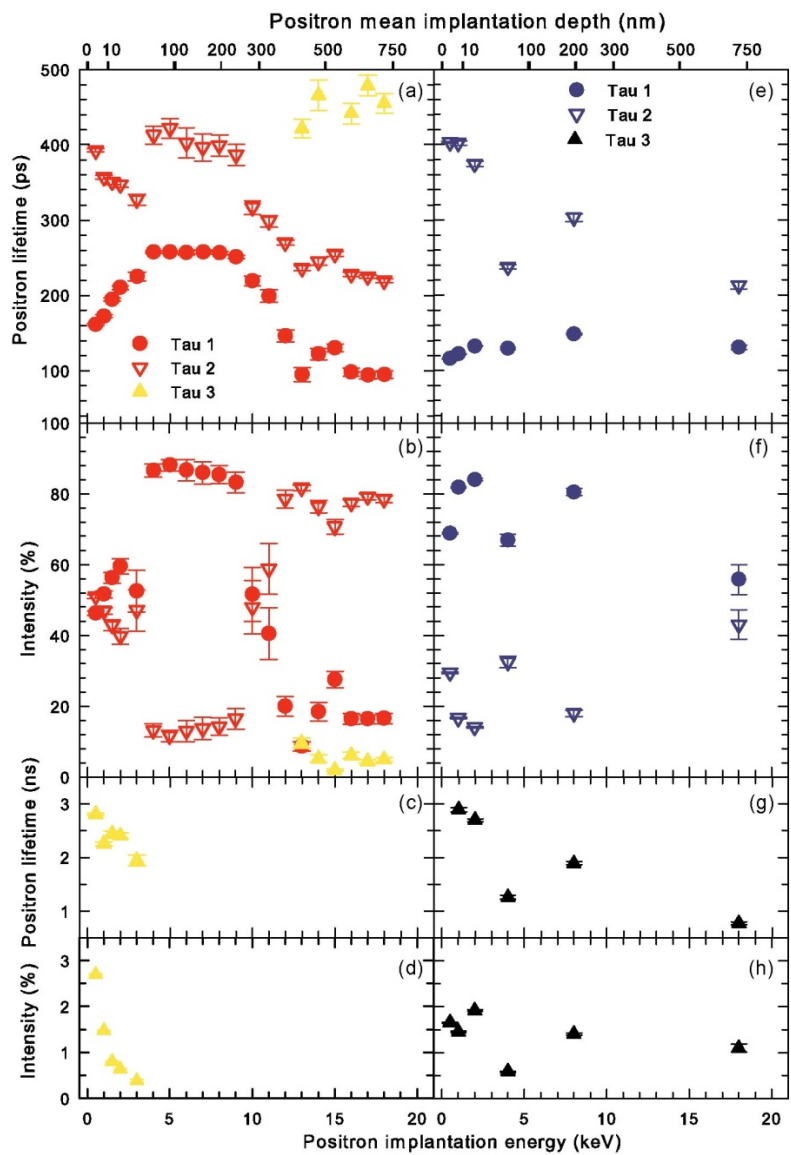